\documentclass[runningheads]{llncs}
\usepackage{multirow}
\usepackage{diagbox}
\usepackage{hyperref}
\usepackage{caption}
\usepackage{subfigure}
\usepackage{booktabs} 
\usepackage{amsmath}
\usepackage{amssymb}    
\usepackage{amsfonts}
\usepackage{diagbox}
\usepackage{graphicx, wrapfig}
\usepackage{array}
\usepackage{xcolor}	
\usepackage{amsthm}    %

\newcommand{\modif}[1]{\textcolor{black}{#1}}

\usepackage{cite}

\begin{document}
\title{Restoring Connectivity in Vascular Segmentations using a Learned Post-Processing Model}
\author{Sophie Carneiro-Esteves\inst{1,2}\orcidID{0009-0000-1098-8030} \and
Antoine Vacavant\inst{1}\orcidID{0000-0001-9616-3282} \and
Odyssée Merveille\inst{2}\orcidID{0000-0002-9918-3761}}

\authorrunning{S. Carneiro-Esteves et al.}
\institute{ Université Clermont Auvergne, CNRS, SIGMA Clermont, Institut Pascal, F-63000 
\email{antoine.vacavant@uca.fr}\\
\and
Univ Lyon, INSA‐Lyon, Université Claude Bernard Lyon 1, UJM-Saint Etienne, CNRS, Inserm, CREATIS UMR 5220, U1294,F‐69100 
\email{\{sophie.carneiro,odyssee.merveille\}@creatis.insa-lyon.fr}\\
}


\maketitle
\vspace{-0.3cm}
\begin{abstract}
Accurate segmentation of vascular networks is essential for computer-aided tools designed to address cardiovascular diseases.
Despite more than thirty years of research, it remains a challenge to obtain vascular segmentation results that preserve the connectivity of the underlying vascular network. Yet connectivity is one of the key features of these tools.
In this work, we propose a post-processing algorithm aiming to reconnect vascular structures that have been disconnected by a segmentation algorithm.
Connectivity being a complex property to model explicitly, we propose to learn this geometric feature either through synthetic data or annotations of the application of interest. The resulting post-processing model can be used on the output of any supervised or unsupervised vascular segmentation algorithm. We show that this post-processing effectively restores the connectivity of vascular networks both in 2D and 3D images, leading to improved overall segmentation results.
\end{abstract}


\section{Introduction}

Blood vessel segmentation is a crucial step for various tasks such as blood flow simulation or 3D modeling,  and enhancing our understanding of vascular networks physiology, and pathologies. However, segmenting blood vessels is challenging due to their thin and tortuous nature, making them easily altered by noise and artifacts. This often results in fragmented blood vessel segmentations which is a major problem for most downstream tasks.

For over thirty years, methods have been introduced to enhance both the quality and the connectivity of blood vessel segmentation. Several unsupervised filtering approaches were first proposed. Vesselness filters~\cite{jonas_article} aims at enhancing the signal from blood vessels and decreasing the one from other non-tubular structures. These filters are designed to detect blood vessels at different scales, employing either a Gaussian-scale paradigm and Hessian-based features extraction ~\cite{frangi1998multiscale, sato1998three}, or a mathematical morphology approach using paths as structuring elements ~\cite{Merveille:PAMI:17}. 
These filters are usually the first step of more complex segmentation pipelines \cite{miraucourt2015variational, merveille2019n, chung2018accurate}. 

However, determining hyperparameters for these filters can be challenging and there is no guarantee on the connectivity of the vascular tree. Alternative unsupervised methods, such as tracking~\cite{carrillo2007recursive} or minimal path methods~\cite{liao2017progressive} can ensure the structure connectivity. Nevertheless, these methods require a time-consuming user interaction to define seed points. 
All these approaches are further limited by having to explicitly model blood vessels.

Supervised methods, and in particular deep learning-based ones, offer the power to represent complex phenomena by learning implicit functions, provided there are sufficient annotations on the target application. Several approaches dedicated to vascular segmentation were proposed \cite{tetteh2020deepvesselnet, mou2021cs2, sanchesa2019cerebrovascular}.
More recently, approaches were developed to improve the vascular connectivity of the segmentation results. Classic vesselness filters were used to help the network model tubular shapes~\cite{shi2022local, peng2023curvilinear}.
Alternative approaches focused on adapting the segmentation architecture to facilitate the learning of a function that preserves connectivity. Attention modules~\cite{vaswani2017attention} were incorporated into architectures such as U-Net~\cite{mou2021cs2}\modif{, and a topology-aware feature synthesis network was proposed to correct the prediction topology based on the Euler characteristic \cite{li2023robust}.}
Proxy tasks were also introduced to help the model focus on the structure topology such as the centerline extraction or distance-map computation \cite{keshwani2020topnet}. 
Many works proposed dedicated loss functions to improve the result connectivity~\cite{shit2021cldice, rougé2023cascaded, lin2023dtu,hu2019topology, clough2020topological, hakim2021regularizer, qiu2023corsegrec}. 
All these connectivity-preserving strategies assume that a large annotated dataset is available, which is rarely the case in vascular imaging applications.

Another research direction was explored consisting of the design of post-processing techniques dedicated to the reconnection of vascular segmentation results. Various algorithms have been suggested, relying on centerlines \cite{du2023retinal}, graphs \cite{mou2020DDN, joshi2011identification}, and contour completion processes \cite{zhang2018reconnection}. These approaches are complex to use due to their high dependence on parameter selection, and none of them provide the code necessary to reproduce or compare their results.

\begin{figure}[!t]
    \centering
    \includegraphics[width=\linewidth]{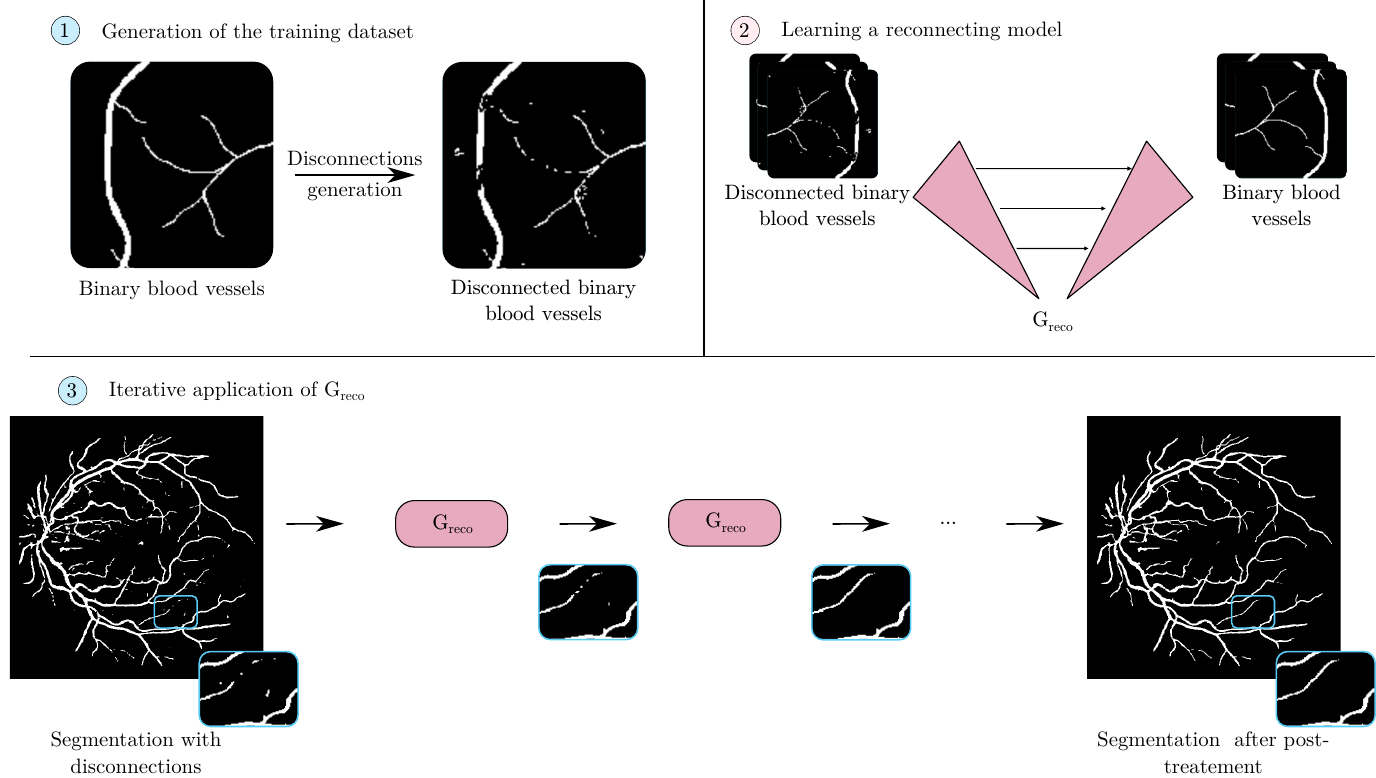}
    \caption{Pipeline of our method. (1) a dataset is generated containing pairs of connected and disconnected vascular structures. (2) this dataset is used to train a model $\text{G}_\text{reco}$ with a residual U-Net architecture. (3) finally, the trained model is iteratively applied on vascular segmentations with disconnections.}
    \label{fig:scheme_framework}
\end{figure}

In ~\cite{esteves4573124plug}, they proposed a strategy to train the reconnecting model and used it to develop an unsupervised plug-and-play segmentation approach. We recognized the potential for the reconnecting term to be applied more broadly as a post-processing step for any type of vascular segmentation result. 
In the present article, we thoroughly investigate this idea. 

In this article, we conduct a comprehensive investigation of this concept. We specifically examine the properties of the reconnecting term, focusing on the impact of the disconnection size parameter and its convergence behavior. Furthermore, we demonstrate the practicality and versatility of this new post-processing strategy by applying it to outputs from various segmentation methods and datasets, and comparing it with a recently developed reconnecting post-processing approach.


\section{Proposed method}\label{sec:method}

In~\cite{esteves4573124plug}, a model, $G_\text{reco}$, based on a residual U-Net~\cite{kerfoot2018left}, is learned to reconnect disconnected vessel-like structures from a binary segmentation result. This model is trained on pairs of images containing connected and disconnected vessel-like structures (see top of Fig.~\ref{fig:scheme_framework}).


The reconnecting term can be learned either based on manual annotations from the dataset of interest, or solely from synthetic images. This approach eliminates the need for an annotated dataset while still producing satisfactory results.
When using synthetic images for training, an algorithm capable of generating random yet realistic disconnections from any binary vascular structure was proposed.
To control the reconnection power of the model, disconnections are generated with various sizes drawn from a Gaussian distribution with mean $s$ and standard deviation $\sigma$.

In this work, we propose to use this reconnecting model as a standalone post-processing step applicable to any binary segmentation result.

Intuitively, the size of the disconnections $s$ of the training dataset should be adjusted to match the disconnection sizes in the segmentation. A small value may be insufficient to reconnect a vessel with a large gap, whereas a large value increases the risk of false reconnections.
To address both small and large disconnections while minimizing false reconnections, we propose applying $G_\text{reco}$ iteratively with a small value of $s$. This approach reconnects vessels gradually rather than attempting long-range reconnections in a single step (see bottom of Fig.~\ref{fig:scheme_framework}). The code of our approach is available at \href{https://github.com/creatis-myriad/plug-and-play-reco-regularization}{https://github.com/creatis-myriad/plug-and-play-reco-regularization}.

%
\section{Experiments}


\subsection{Experimental set up}

To analyse and demonstrate the effectiveness of our method, we tested our framework in both 2D and 3D and used both synthetic and real datasets.
In 2D, we used the DRIVE~\cite{staal2004ridge} dataset composed of 40 retinophotographies and their manual vascular annotations, and the STARE ~\cite{hoover2000locating} dataset, which includes 20 annotated retinophotographies. The STARE dataset was used to train our reconnecting model, while the DRIVE dataset served as the test set for applying $G_\text{reco}$. Additionnaly, we generated a synthetic dataset comprising 20 synthetic vascular trees with OpenCCO~\cite{OpenCCO}, which was also employed as a training dataset for $G_\text{reco}$.
In 3D, we used the Bullitt\cite{bullitt2005vessel} and IXI\footnote{\href{https://brain-development.org/ixi-dataset/}{https://brain-development.org/ixi-dataset/}}~\cite{Valderrama2023} datasets composed of $33$ and $22$ brain magnetic resonance angiographies (MRA) respectively and their manual vascular annotations. IXI served to train our reconnecting model while Bullitt was used as a test set to apply $G_\text{reco}$.

The disconnection algorithm was used on the DRIVE, STARE and IXI datasets to generate disconnected vascular trees with several mean disconnection sizes ($s\in {6, 8, 10, 12}$). We experimentally set the standard deviation $\sigma=4$ for DRIVE and STARE and to $\sigma=2$ for IXI.

The backbone architecture for $G_\text{reco}$ is a residual U-Net model \cite{kerfoot2018left} trained for $1000$ epochs in 2D and $3000$ epochs in 3D. We used an Adam optimizer with a learning rate of $10^{-3}$. We employed a weighted Dice loss function as presented in~\cite{esteves4573124plug} and a batch size of $32$ in 2D and $4$ in 3D. In 2D, the models were trained with an 80\% split for training and 20\% for validation, while in 3D, the split was 90\% for training and 10\% for validation. The final model was selected based on the best validation loss achieved during the training.


We selected three metrics to evaluate the segmentations. The classic Dice coefficient (DSC) assesses the overall quality of the segmentation. The Average Symmetric Surface Distance (ASSD) evaluates the segmentation geometry by measuring the distances between each element of the segmentation contour and the corresponding contour in the annotation. Finally, the error ratio of the number of connected components $\epsilon_{\beta_0}$ evaluates the segmentation connectivity. This error ratio is defined as $\epsilon_{\beta_0} = \left| \frac{{\beta_0 - {\beta_0}_{\text{gt}}}}{{{\beta_0}_{\text{gt}}}} \right|$, 
with $\beta_0$ the number of connected components of the segmentation and  ${\beta_0}_{\text{gt}}$ the number of connected component of the annotation. The error ratio was preferred over the value of $\beta_0$ as $\beta_0$ is usually larger than 1 in the DRIVE and Bullitt annotations. 

\subsection{Ablation study}

\begin{figure}[tb]
    \centering
    \subfigure[Annotation]{\includegraphics[width=0.30\linewidth]{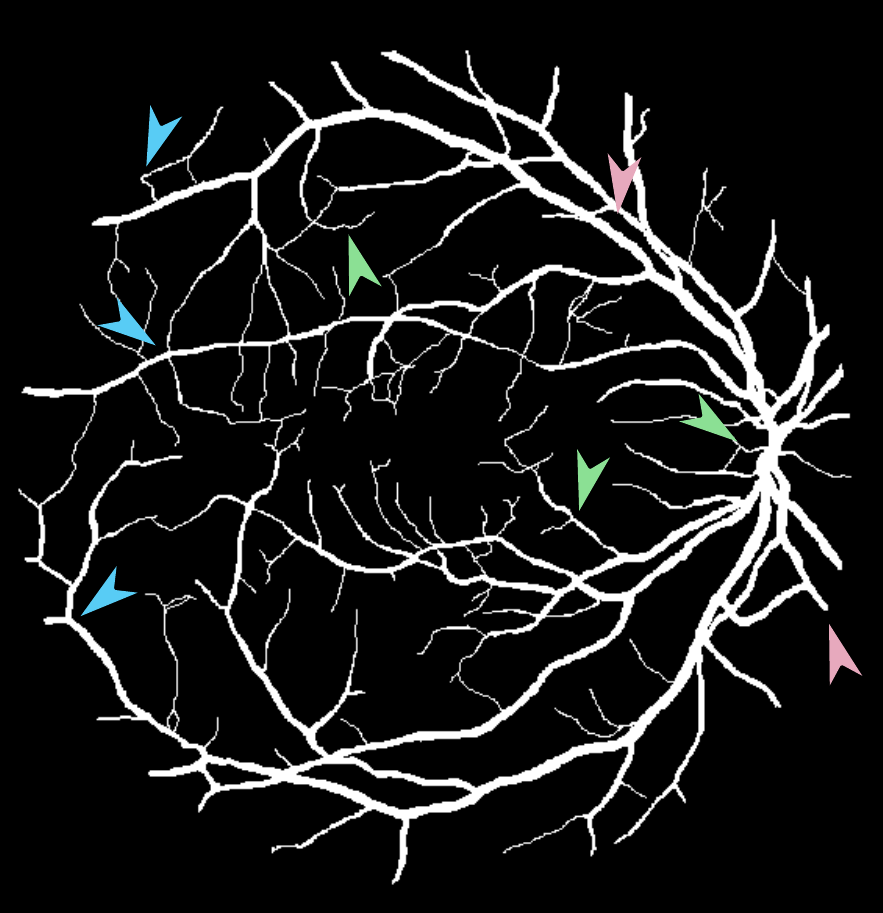}}
    \subfigure[Disconnected annotation with $s=12$]{\includegraphics[width=0.30\linewidth]{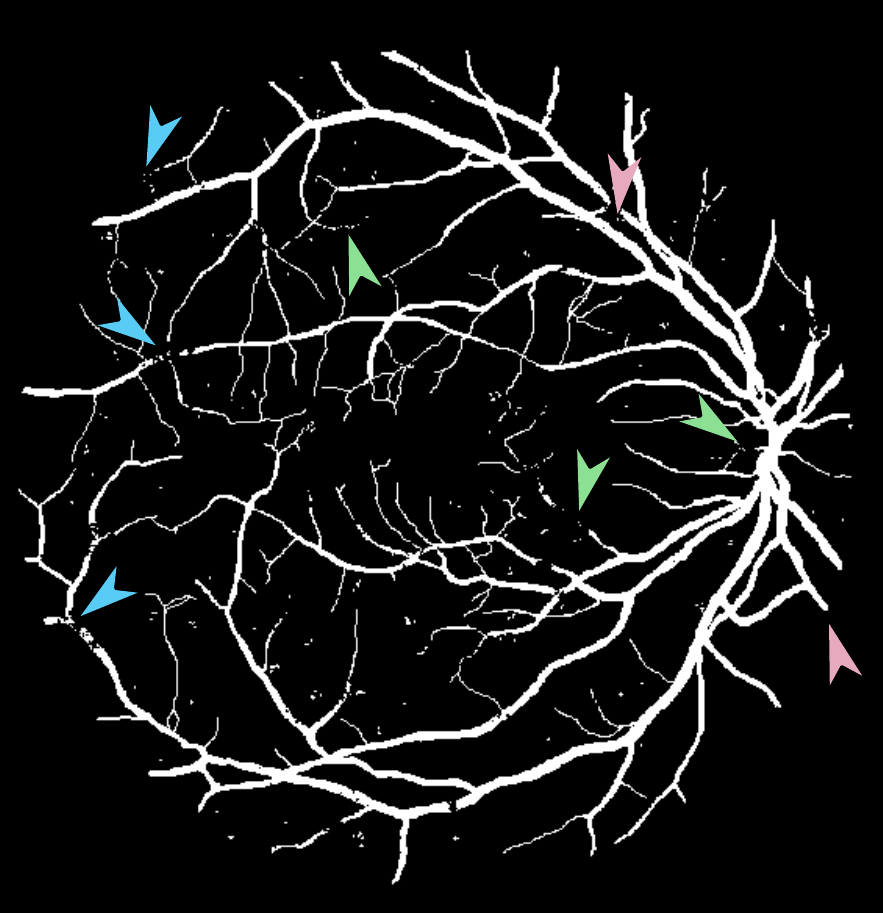}}
    \subfigure[$\text{G}_{\text{reco}, s = 6}$]{\includegraphics[width=0.30\linewidth]{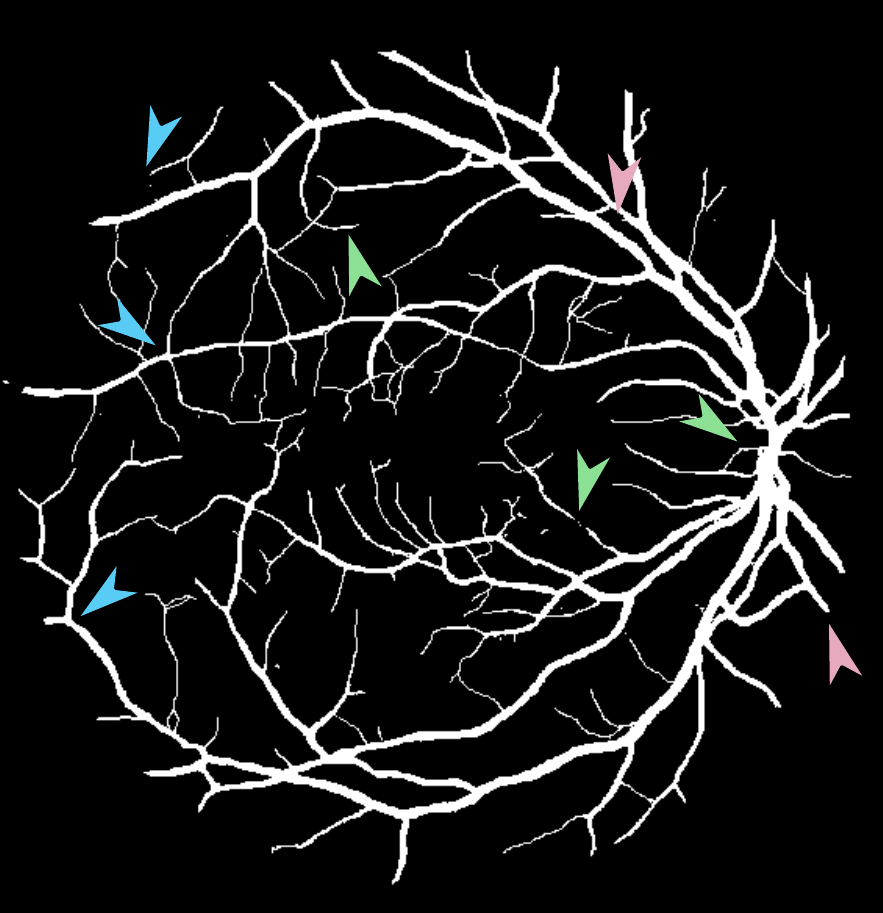}}\\
    \subfigure[$\text{G}_{\text{reco}, s = 8}$]{\includegraphics[width=0.30\linewidth]{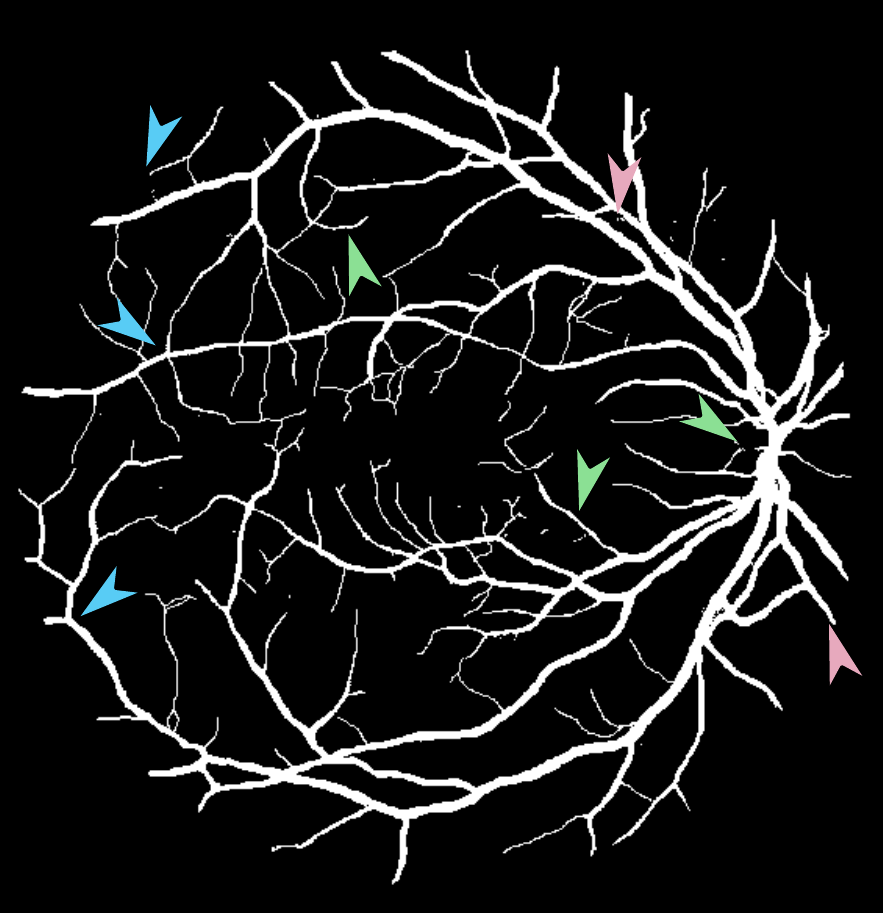}}
    \subfigure[$\text{G}_{\text{reco}, s = 10}$]{\includegraphics[width=0.30\linewidth]{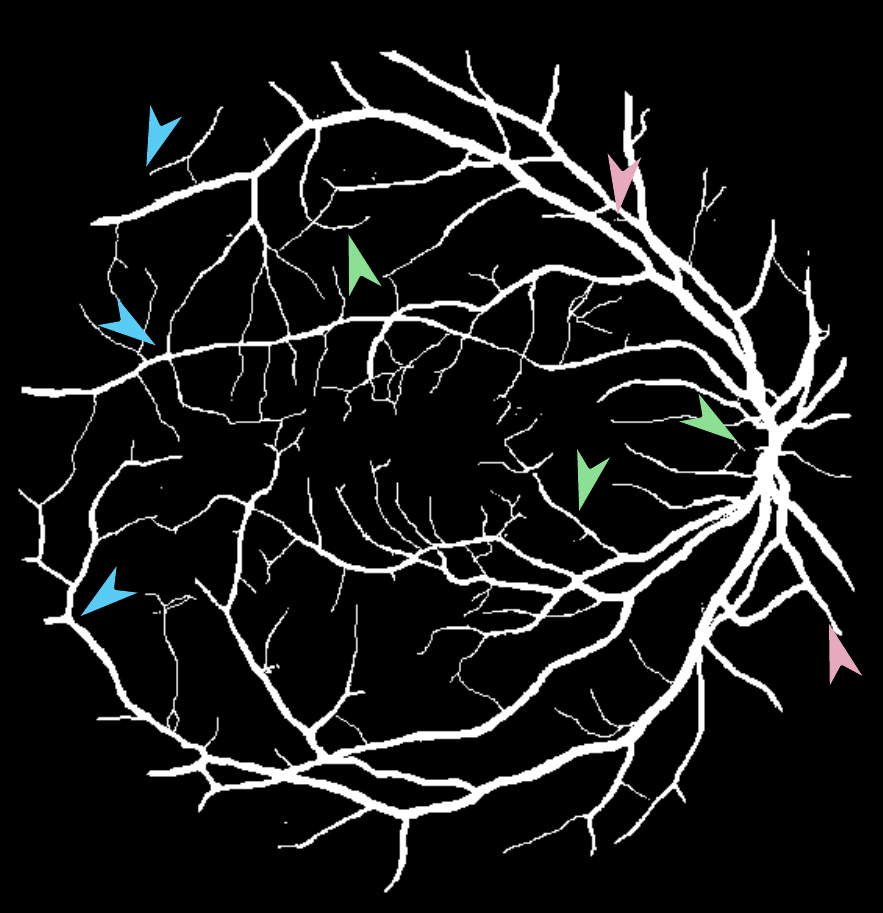}}
    \subfigure[$\text{G}_{\text{reco}, s = 12}$]{\includegraphics[width=0.30\linewidth]{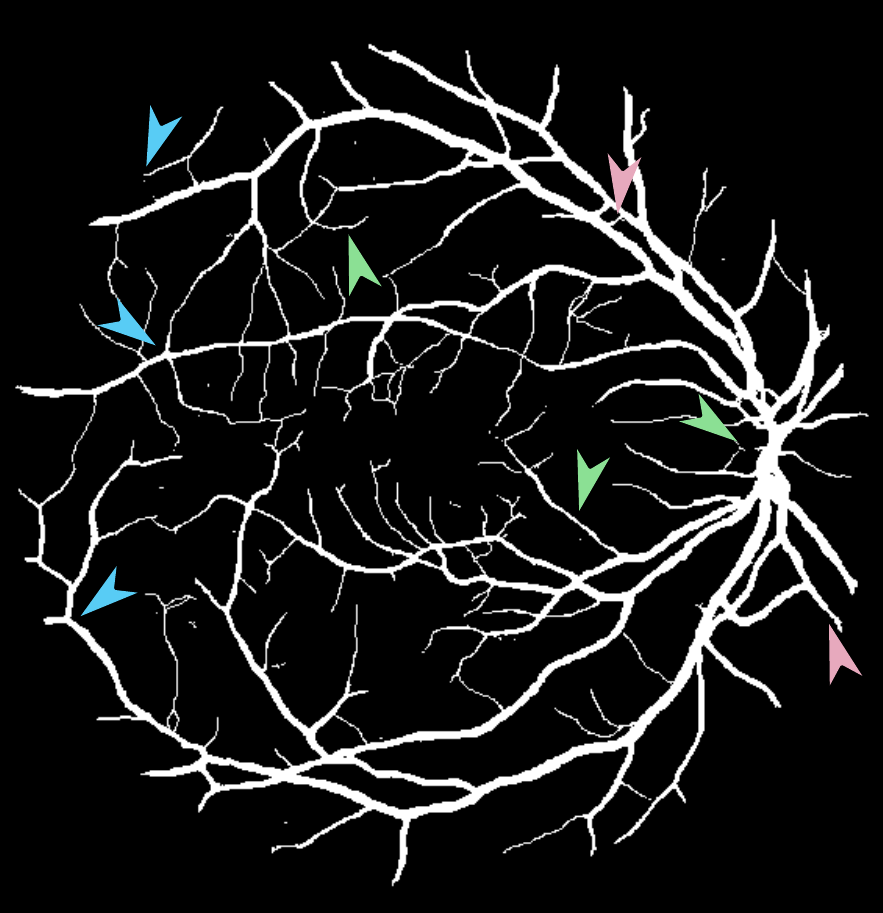}}
    \caption{Qualitative results of the influence of the disconnection size experiment. $G_\text{reco}$ was trained with a disconnection size $X$ (with $X\in {6, 8, 10, 12}$), denoted $G_{\text{reco},s=X}$, on the OpenCCO dataset. A drive dataset annotation (a) was disconnected with $s=12$ to obtain (b), followed by a single application of $G_{\text{reco},s=X}$ on (b), resulting in (c-f). Blue arrows indicate successful reconnections, pink arrows highlight incorrect reconnections, and green arrows denote differences between the results.}
    \label{appendix:influence_size}
\end{figure}

\subsubsection{Influence of the size of the disconnections}\label{subsubsec: deconnection size}

The training dataset is a key element of our framework as it defines the concept of reconnection. The main parameter of this training dataset is the mean size of a disconnection, denoted as $s$.
We trained four different models, denoted $G_{\text{reco},s=X}$ (with $X\in {6, 8, 10, 12}$), on the OpenCCO dataset which has been disconnected with a mean disconnection size $X$.
We then tested these $4$ models on the annotations of the DRIVE dataset that have also been disconnected with increasing values of $s$ ($s=6, s=8, s=10$ or $s=12$). Hence each model will be tested on disconnection sizes it has not been trained for. In this experiment, $G_\text{reco}$ has only been applied once to the segmentation results.
The results are presented in Table~\ref{tab:result_1_iteration}.

As expected, the Dice coefficient of all models does not show a significant improvement after applying our post-processing, as the reconnection fragments represent only a small portion of the overall vessels. However both $\epsilon_{\beta_0}$ and $ASSD$ significantly decrease compared to the initial segmentation (Before $G_\text{reco})$, indicating that our post-processing successfully reconnected fragments of vessels. 

We also observe a correlation between the size of the disconnections in the training dataset and the reconnections that occur. Models trained on larger disconnections tend to perform better on large disconnections (see Figure.~\ref{appendix:influence_size}), while still facing challenges in reconnecting smaller disconnections. 
The model trained with disconnections with a size of $s = 8$, seems a good compromise, effectively reconnecting vessels while minimizing excessive false connections. 

\begin{table}[t]
\centering
\scalebox{0.65}{
\begin{tabular}{cccccccccccccccc}
 & \multicolumn{3}{c}{$s = 6$} &  & \multicolumn{3}{c}{$s = 8$} &  & \multicolumn{3}{c}{$s = 10$}& & \multicolumn{3}{c}{$s = 12$} \\ \cline{2-4} \cline{6-8} \cline{10-12} \cline{14-16}
\multirow{-2}{*}{\diagbox[width=10em]{Training}{Test}} & DSC $\uparrow$ & ASSD $\downarrow$ & $\epsilon_{\beta_0}$ $\downarrow$&  & DSC $\uparrow$& ASSD $\downarrow$& $\epsilon_{\beta_0}$ $\downarrow$&  & DSC $\uparrow$& ASSD $\downarrow$& $\epsilon_{\beta_0}$ $\downarrow$&  & DSC $\uparrow$& ASSD $\downarrow$& $\epsilon_{\beta_0} \downarrow$ \\ \hline
\multicolumn{1}{c}{} &0.979 & 0.202 & 96.811 &&0.974 & 0.22 & 107.367 &&0.97 & 0.232 & 122.198 &&0.963 & 0.243 & 132.617 \\
\multicolumn{1}{c}{\multirow{-2}{*}{Before $\text{G}_{\text{reco}}$}} & $\pm$ 0.004 &  $\pm$ 0.06 &  $\pm$ 67.065 & & $\pm$ 0.004 &  $\pm$ 0.071 &  $\pm$ 71.883 & & $\pm$ 0.005 &  $\pm$ 0.057 &  $\pm$ 83.665 & & $\pm$ 0.007 &  $\pm$ 0.054 &  $\pm$ 86.553 \\ \hline \hline
\multicolumn{1}{c}{} &0.983 & 0.074 & \textbf{11.429} &&0.98 & 0.085 & \textbf{14.095} &&0.978 & 0.101 & \textbf{15.485} &&0.974 & 0.116 & \textbf{17.619} \\
\multicolumn{1}{c}{\multirow{-2}{*}{$G_{\text{reco}, s = 6}$}} & $\pm$ 0.003 &  $\pm$ 0.012 &  $\pm$ 8.783 & & $\pm$ 0.003 &  $\pm$ 0.015 &  $\pm$ 10.92 & & $\pm$ 0.004 &  $\pm$ 0.019 &  $\pm$ 10.758 & & $\pm$ 0.004 &  $\pm$ 0.022 &  $\pm$ 13.8 \\
\multicolumn{1}{c}{} &\textbf{0.985} & \textbf{0.067} & 15.461 &&\textbf{0.983} & \textbf{0.077} & 17.301 &&\textbf{0.981} &\textbf{ 0.09} & 19.365 &&\textbf{0.977} & \textbf{0.103} & 23.039 \\
\multicolumn{1}{c}{\multirow{-2}{*}{$G_{\text{reco}, s = 8}$}} & $\pm$ 0.002 &  $\pm$ 0.013 &  $\pm$ 11.256 & & $\pm$ 0.003 &  $\pm$ 0.014 &  $\pm$ 12.69 & & $\pm$ 0.004 &  $\pm$ 0.019 &  $\pm$ 14.269 & & $\pm$ 0.004 &  $\pm$ 0.02 &  $\pm$ 18.927 \\
\multicolumn{1}{c}{} &0.984 & 0.078 & 14.61 &&0.981 & 0.089 & 16.829 &&0.979 & 0.101 & 18.82 &&0.975 & 0.117 & 22.056 \\
\multicolumn{1}{c}{\multirow{-2}{*}{$G_{\text{reco}, s = 10}$}} & $\pm$ 0.003 &  $\pm$ 0.015 &  $\pm$ 11.223 & & $\pm$ 0.003 &  $\pm$ 0.016 &  $\pm$ 13.051 & & $\pm$ 0.004 &  $\pm$ 0.023 &  $\pm$ 14.516 & & $\pm$ 0.004 &  $\pm$ 0.017 &  $\pm$ 17.864 \\
\multicolumn{1}{c}{} &0.982 & 0.089 & 16.232 &&0.98 & 0.102 & 18.561 &&0.977 & 0.118 & 19.776 &&0.974 & 0.126 & 20.612 \\
\multicolumn{1}{c}{\multirow{-2}{*}{$G_{\text{reco}, s = 12}$}} & $\pm$ 0.003 &  $\pm$ 0.015 &  $\pm$ 12.213 & & $\pm$ 0.003 &  $\pm$ 0.017 &  $\pm$ 13.75 & & $\pm$ 0.004 &  $\pm$ 0.024 &  $\pm$ 14.034 & & $\pm$ 0.004 &  $\pm$ 0.019 &  $\pm$ 16.189\\[1em]
\end{tabular}}

\caption{Results of applying $G_\text{reco}$ trained on the OpenCCO dataset with several values of $s$, and applied on the Drive dataset with several values of $s$.}
\label{tab:result_1_iteration}
\end{table}

\subsubsection{Convergence of the proposed approach \label{subsec: convergence}}

In the previous experiment, we evaluated our models after a single application to better understand their behavior. However, our goal is to use $G_\text{reco}$ iteratively to address disconnections that can not be fully reconnected in a single iteration. In the this experiment, we analyze the interest of this iterative approach. 
Additionally, given the limited number of disconnections in an image, we anticipate that applying $G_\text{reco}$ iteratively will converge to a fixed-point image where all disconnections are filled. Thus, we also explore the experimental convergence of our framework in this experiment.

\begin{figure}[tb]
    \centering
    \subfigure{\includegraphics[width=0.45\linewidth]{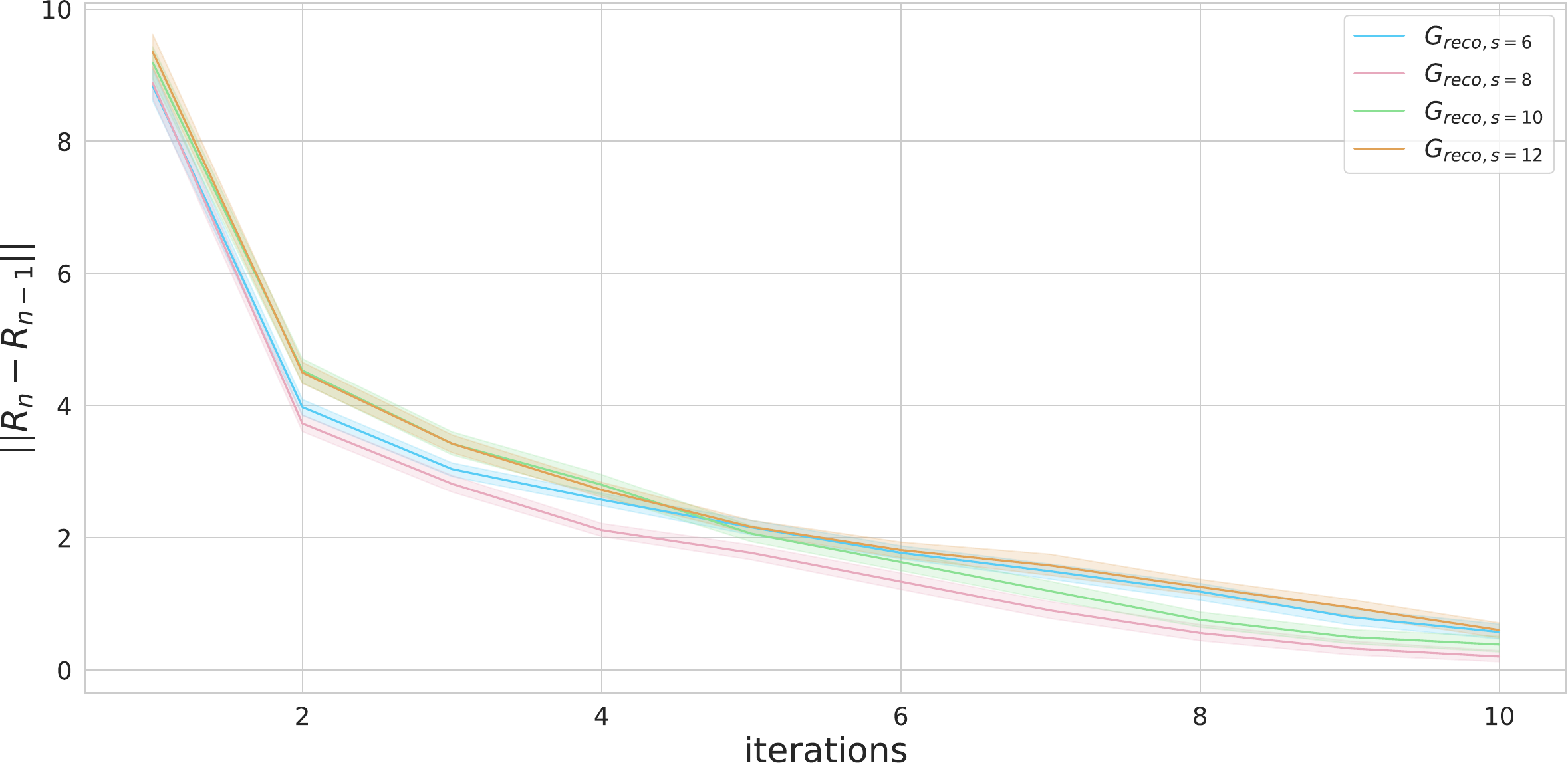}}
    \subfigure{\includegraphics[width=0.45\linewidth]{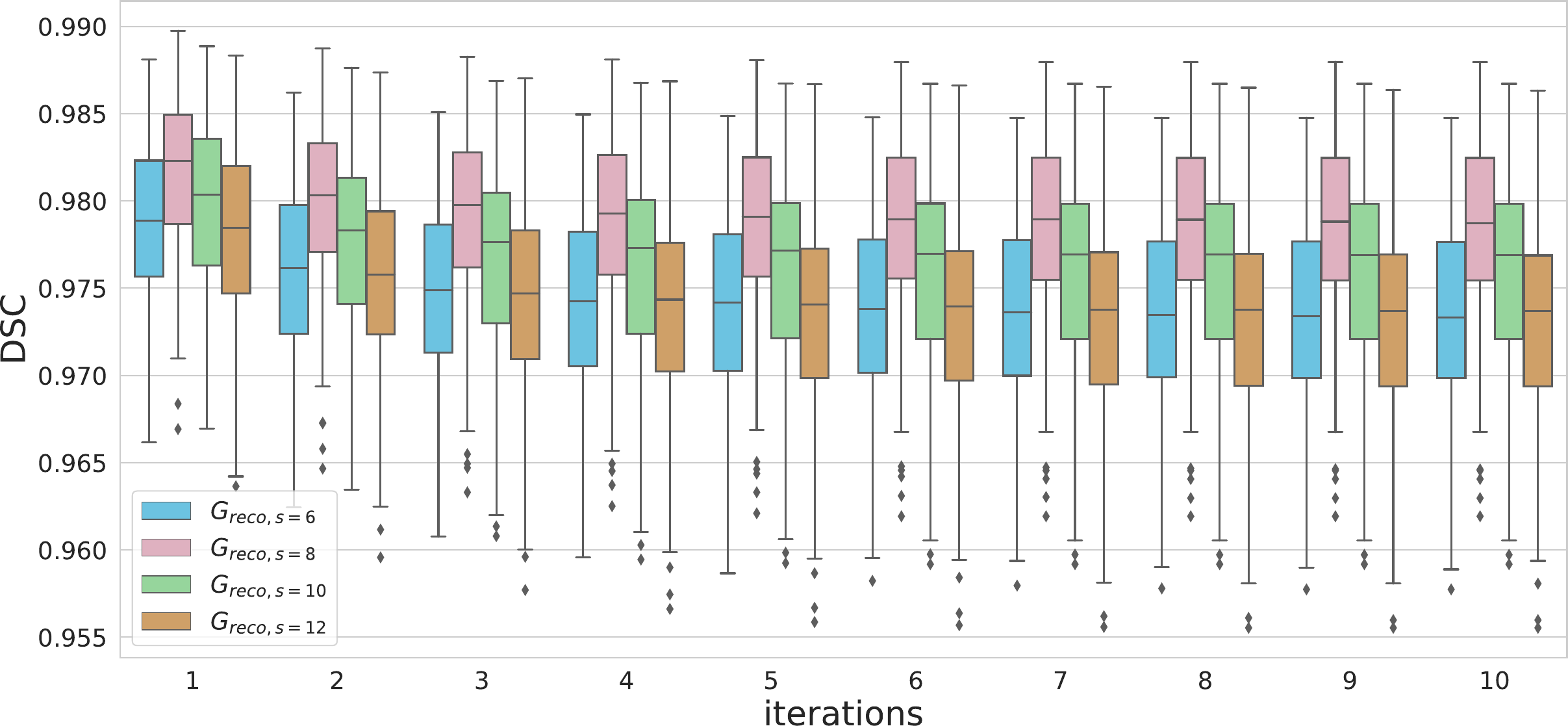}}\\
    \subfigure{\includegraphics[width=0.45\linewidth]{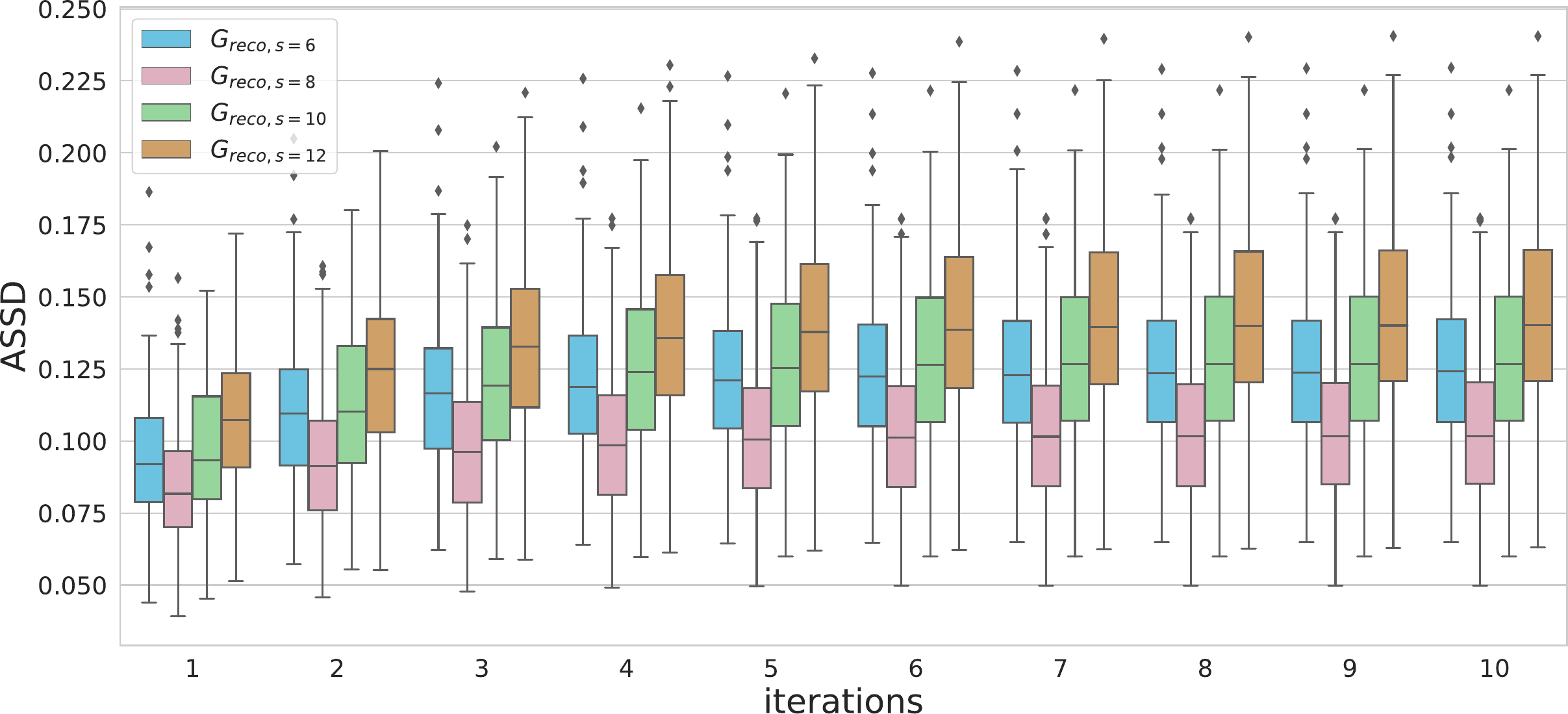}}
    \subfigure{\includegraphics[width=0.45\linewidth]{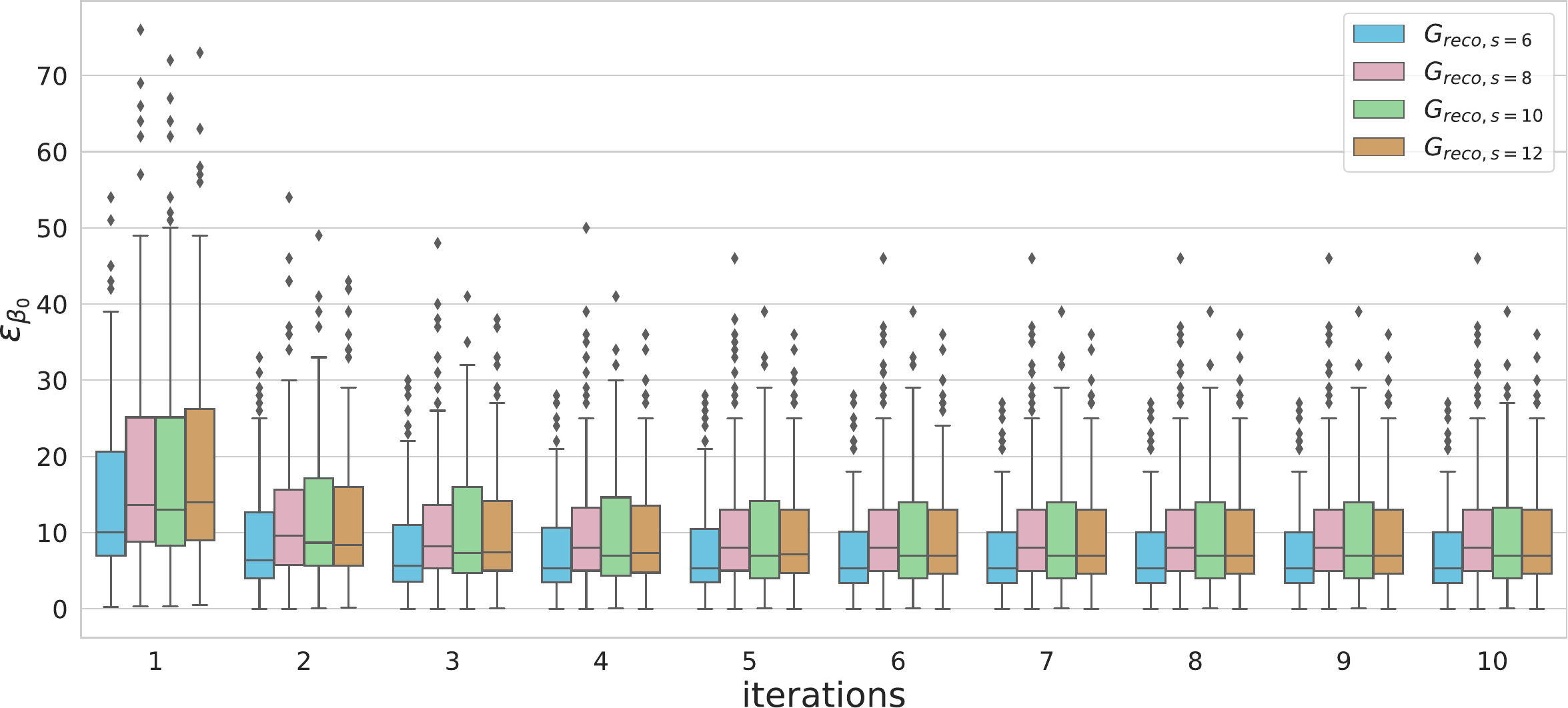}}

    \caption{Results of applying $\text{G}_{\text{reco}}$, trained on OpenCCO, on 
    the DRIVE annotations artificially disconnected with sizes $s\in\{6, 8, 10, 12\}$. (a) convergence curves displaying the $\ell_2$ norm of the difference of the last two consecutive results. (b-d) quantitative results of our models applied with an increasing number of iterations.}
    \label{fig:convergence}
\end{figure}

\begin{figure}[tb]
    \centering
    \includegraphics[width=\linewidth]{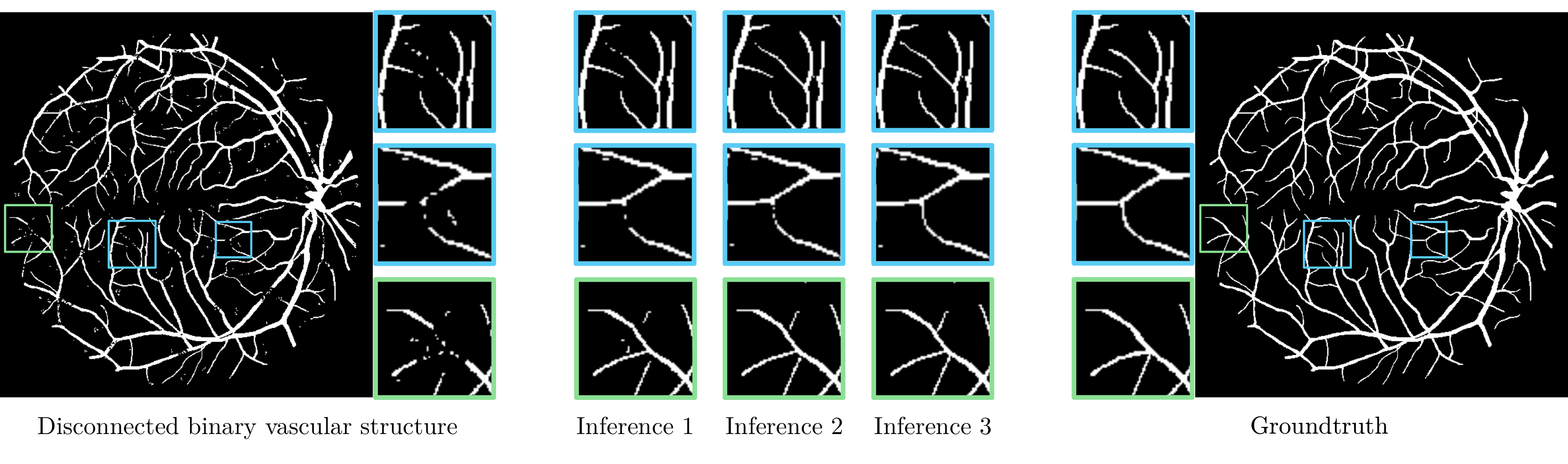}
    \caption{Evolution of a DRIVE disconnected manual annotation (with $s=12$) through successive application of $\text{G}_{\text{reco}, s=8}$.}
    \label{appendix:evolution} 
\end{figure}

We used the same $4$ reconnecting models $G_{\text{reco},s=X}$ ($X \in {6, 8, 10, 12}$) and applied each one on the Drive dataset which has been disconnected with several mean disconnection sizes ($s \in {6, 8, 10, 12}$). The results are presented in Figure~\ref{fig:convergence} and in Figure~\ref{appendix:evolution}.

We observe that the iterative application of $G_\text{reco}$ converges as shown in Figure~\ref{fig:convergence} (a). Interestingly this convergence occurs even though $G_\text{reco}$ was applied to images with disconnection sizes different from those used in the training set. This highlights the robustness of our approach.
In the blue boxes of Figure~\ref{appendix:evolution}, we can observe that some disconnections, not successfully reconnected with a single iteration, are gradually reconnected with additional iterations.
However, it is worth noting that some false reconnections may still occur, as shown by the green boxes. Interestingly, these false reconnections often occur when noise fragments are present near a vessel, and $G_\text{reco}$ uses them to create realistic vessels.



Applying our method on artificially disconnected images let us analyze precisely the behavior of our reconnecting term but this introduced a bias since we used the same algorithm to generate disconnections in both the training and test datasets. The next section will explore the behavior of our reconnecting approach in a more realistic scenario.

\subsection{Applications}
In this experiment, we applied our post-processing to real 2D and 3D segmentation results obtained from both an unsupervised variational approach and a supervised deep learning approach. Specifically, we used the variational segmentation method proposed by Chan \textit{et al.}~\cite{chan2006algorithms} with a total variation~\cite{rudin1992nonlinear} regularization.
A gold standard U-Net architecture was used as the supervised approach as detailed in \cite{esteves4573124plug}. 
The variational segmentation yields disconnected and noisy results whereas the supervised one produces more complete and connected segmentations. We chose to evaluate our framework on these two different types of results to highlight the versatility of our post-processing method.

In 2D, we trained our reconnecting model $G_\text{reco}$ either on the synthetic OpenCCO dataset, denoted as $G_{\text{reco}, CCO}$, or the STARE dataset, denoted as $G_{\text{reco}, STARE}$, that have both been disconnected with a mean disconnection size set to $s=8$. We ran both segmentation strategies (variational and deep-learning) on the DRIVE dataset and applied our post-processing. 

In 3D, we trained our reconnecting model on the IXI dataset, denoted as $G_{\text{reco},{\text{IXI}}}$, that have been disconnected with a mean disconnection size set to $s=8$. We ran both segmentation strategies on the Bullitt dataset and applied our post-processing.

\begin{figure}[tb]
    \centering
    \subfigure[Annotation]{\includegraphics[width=0.19\linewidth]{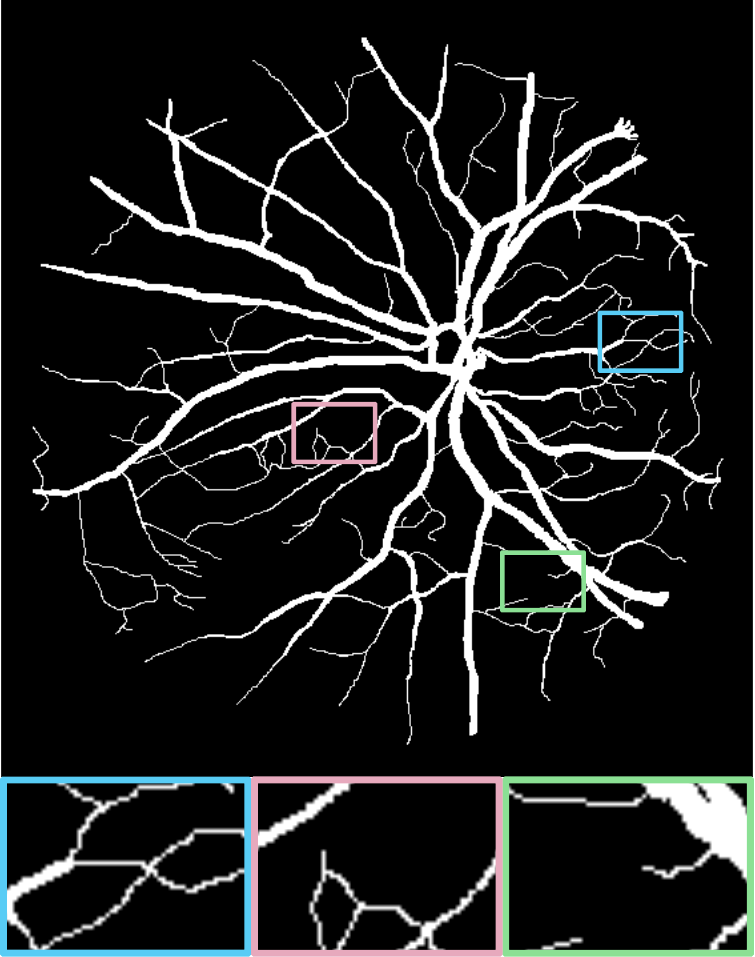}} 
    \subfigure[\scalebox{0.9}{Segmentation}]{\includegraphics[width=0.19\linewidth]{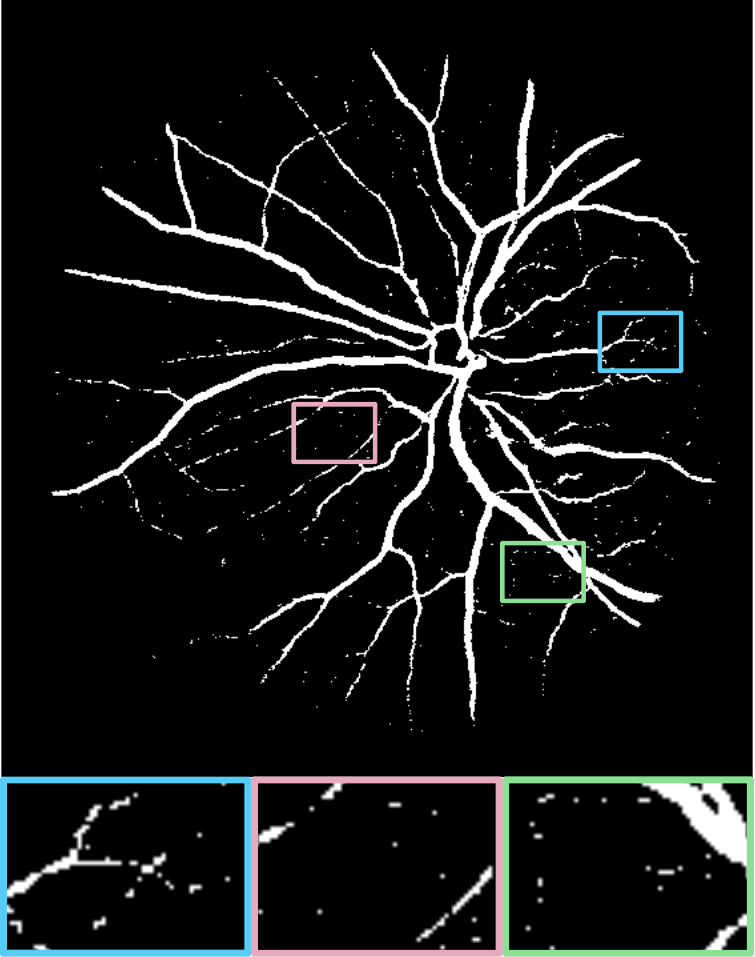}} 
    \subfigure[PRW~\cite{mou2020DDN}]{\includegraphics[width=0.19\linewidth]{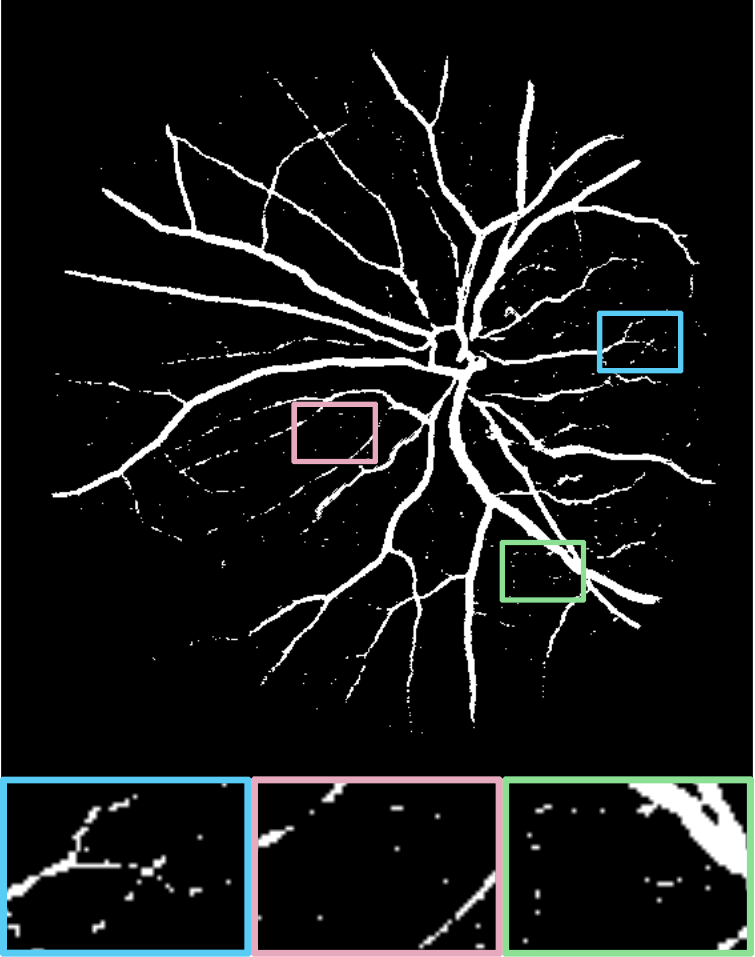}} 
    \subfigure[${\text{G}_{\text{reco},\text{CCO}}}$]{\includegraphics[width=0.19\linewidth]{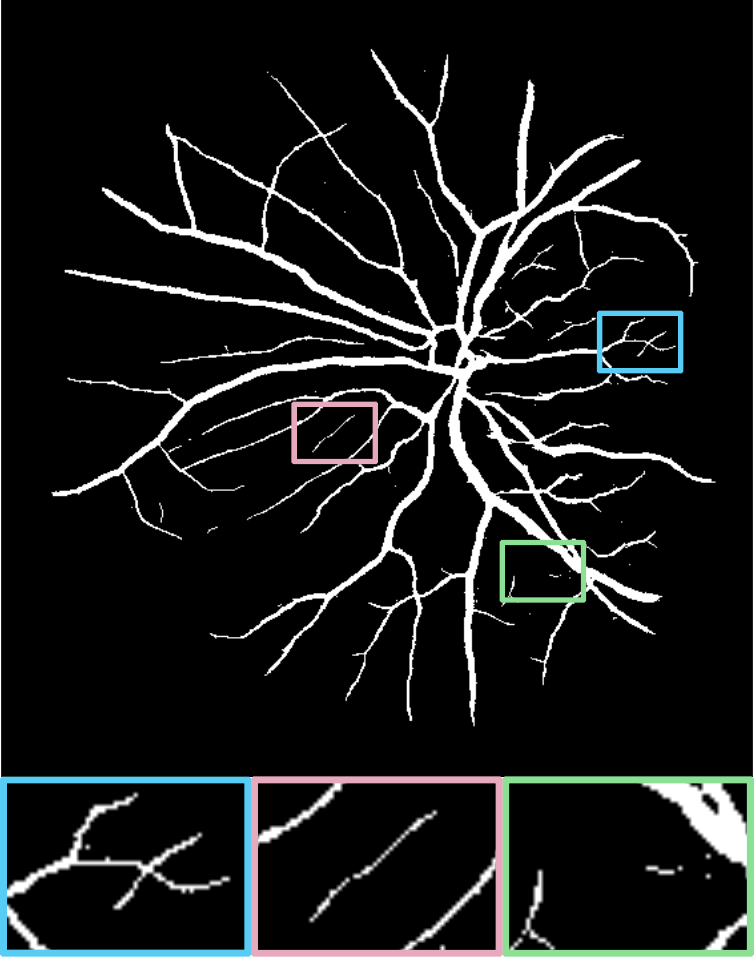}} 
    \subfigure[${\text{G}_{\text{reco},\text{STARE}}}$]{\includegraphics[width=0.19\linewidth]{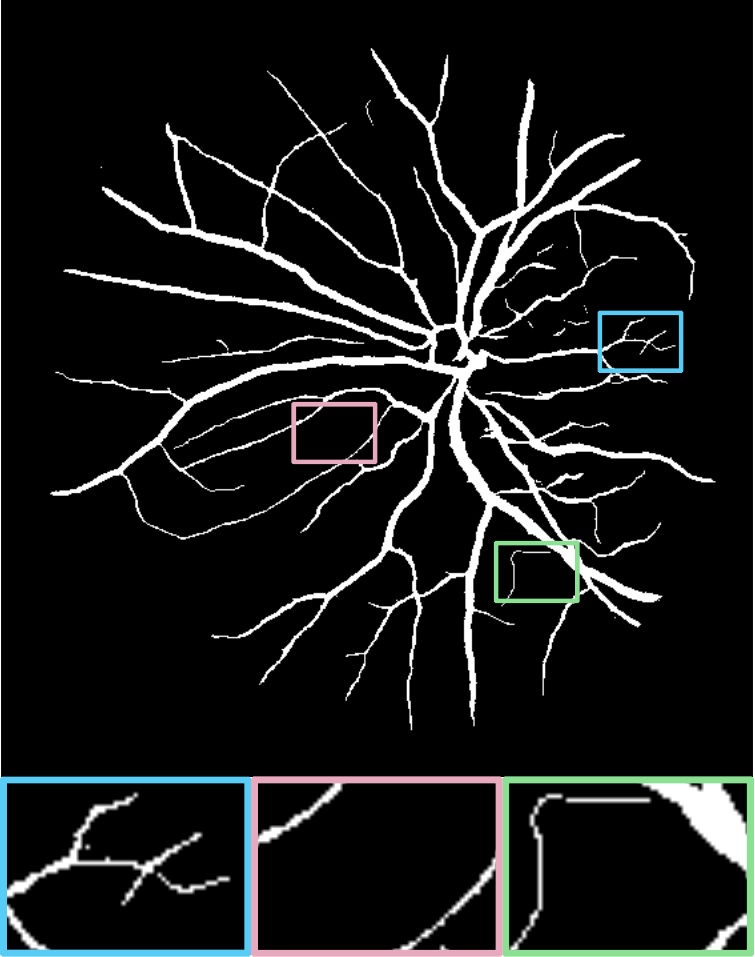}} 
    \\
    \subfigure[Annotation]{\includegraphics[width=0.19\linewidth]{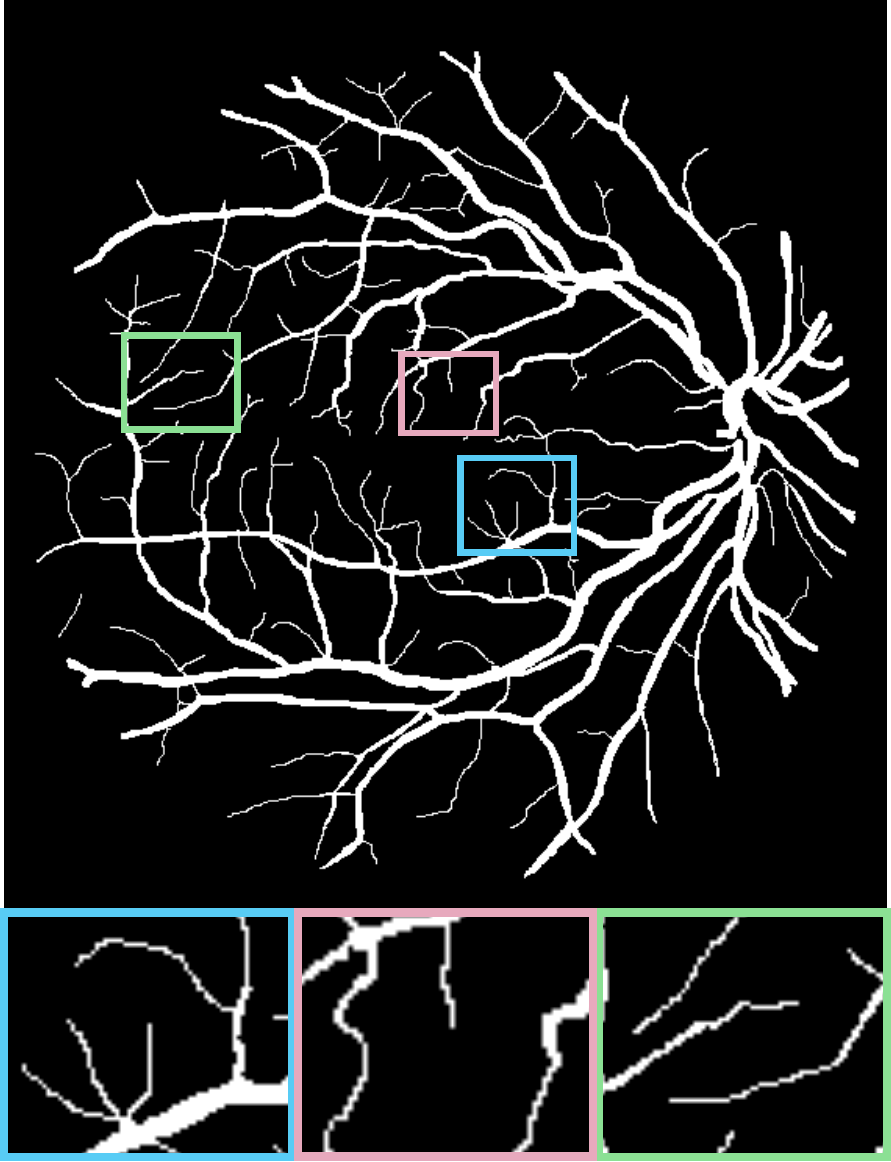}} 
    \subfigure[\scalebox{0.9}{Segmentation}]{\includegraphics[width=0.19\linewidth]{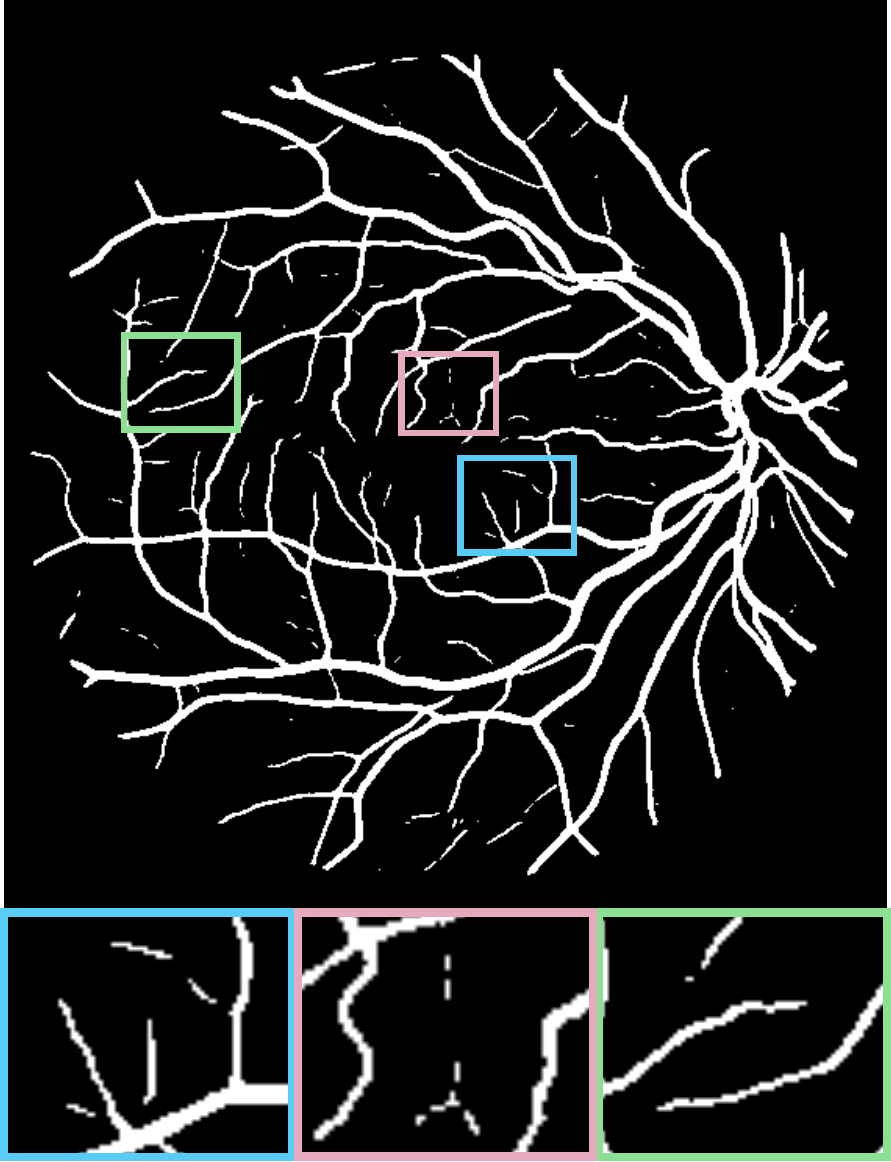}} 
    \subfigure[PRW~\cite{mou2020DDN}]{\includegraphics[width=0.19\linewidth]{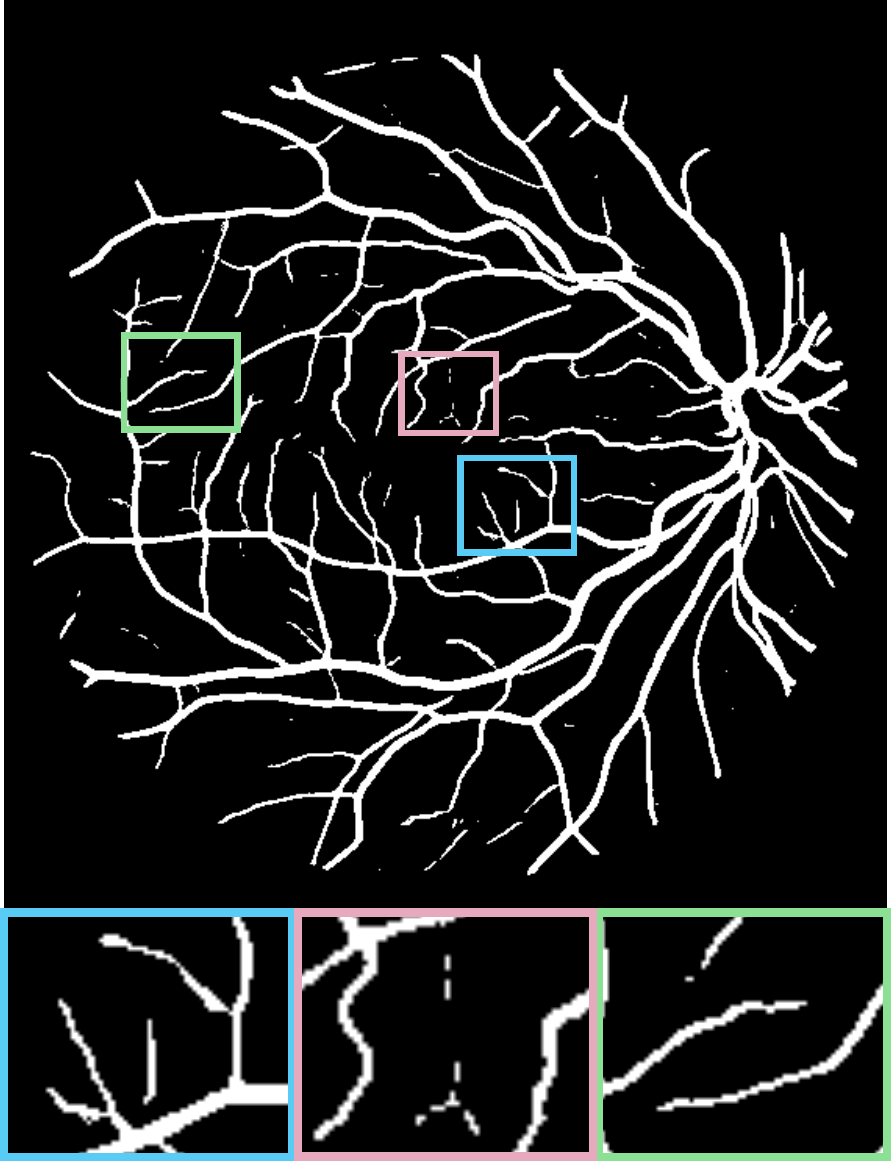}} 
    \subfigure[${\text{G}_{\text{reco},\text{CCO}}}$]{\includegraphics[width=0.19\linewidth]{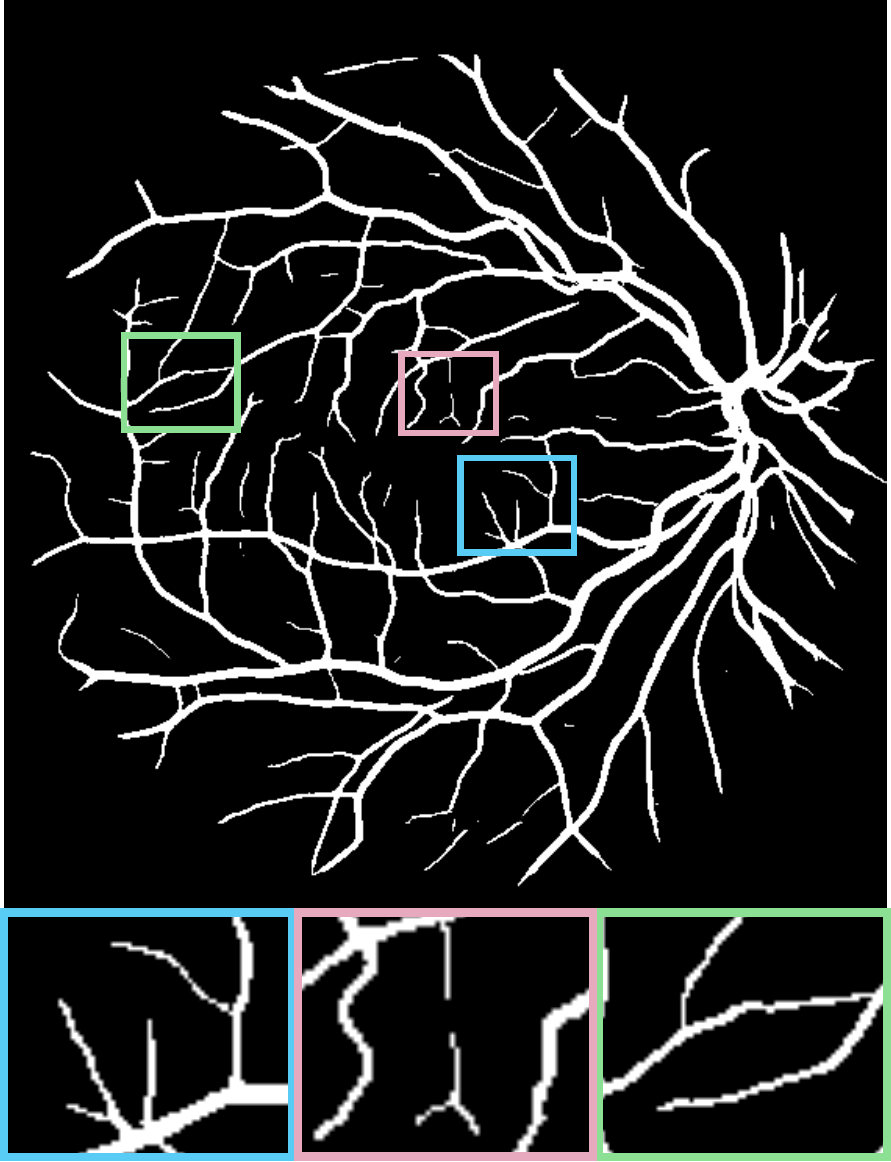}} 
    \subfigure[${\text{G}_{\text{reco},\text{STARE}}}$]{\includegraphics[width=0.19\linewidth]{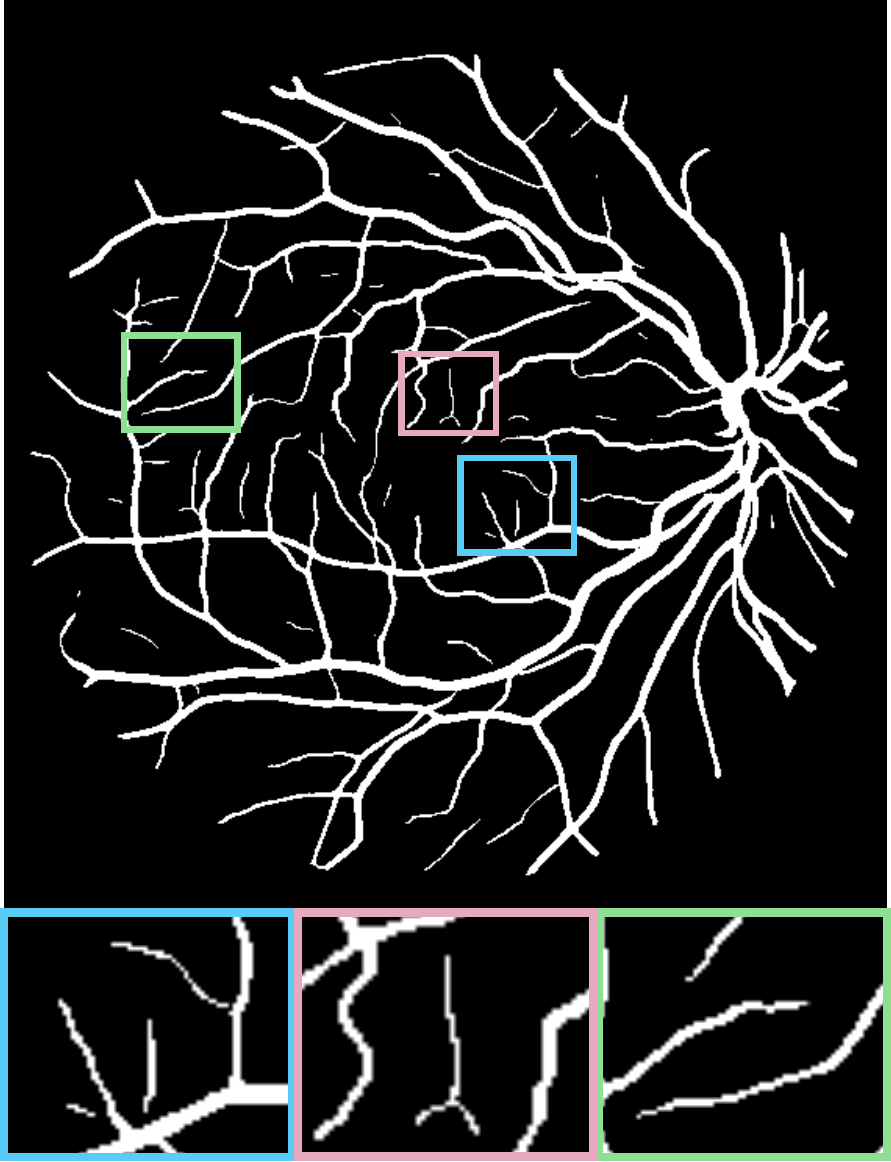}} 
    \caption{Results of our post-processing applied to a DRIVE segmentation result from the variational (top row) and deep learning approach (bottom row).
    (b) and (g) depict segmentation results from variational and deep learning approaches respectively, before post-processing with $G_\text{reco}$. $G_\text{reco}$ is trained on either the synthetic OpenCCO dataset (d) and (i) or the real STARE dataset (e) and (j). (c) and (h) represent the post-processing proposed in ~\cite{mou2020DDN}.}
    \label{fig:results_2D}
\end{figure}

As previously mentioned, to the best of our knowledge, no other reconnecting post-processing approach provides open-access code. Despite this, for comparative purposes, we implemented the probability regularized walk (PRW) algorithm, as proposed in~\cite{mou2020DDN}, and conducted a comparison of our results with it.
This method integrates the probability output of a neural network and the directions of broken vessel segments into a regularized walk algorithm to reconnect them to the main component. The hyperparameters of this method have been optimized for each image using a grid search. Due to incomplete details in the original article~\cite{mou2020DDN}, we made certain choices that might differ from those of the authors. These decisions were made to ensure a fair comparison. Our implementation can be accessed at \href{https://github.com/creatis-myriad/plug-and-play-reco-regularization}{https://github.com/creatis-myriad/plug-and-play-reco-regularization}.
%
Results are summarized in Table~\ref{tab:chap3_variationnelle}, Figure~\ref{fig:results_2D}.
\begin{table}[tb]
\centering
\scalebox{0.8}{
\begin{tabular}{ccccccccc}
&& \multicolumn{3}{c}{Variational approach} &  & \multicolumn{3}{c}{Deep Learning} \\ \cline{3-5} \cline{7-9}
&\multicolumn{-1}{c}{\multirow{-2}{*}{\diagbox[width=10em]{training}{Test}}} & DSC & ASSD & $\epsilon_{\beta_0}$ &  & DSC & ASSD & $\epsilon_{\beta_0}$  \\ \hline
\multicolumn{1}{c|}{}&\multicolumn{1}{c}{} &  0.758 & 2.017 & 98.543 & & \textbf{0.811} & \textbf{1.155} & 34.037 \\
\multicolumn{1}{c|}{}& \multirow{-2}{*}{Segmentation}& $\pm$ 0.025 & $\pm$ 0.452 &  $\pm$ 91.876& & $\pm$ 0.015 & $\pm$ 0.181 &  $\pm$ 26.858 \\
\multicolumn{1}{c|}{}&\multicolumn{1}{c}{} & 0.758 & \textbf{2.012} & 88.909 & & 0.808 & 1.171 & 22.017 \\
\multicolumn{1}{c|}{}& \multirow{-2}{*}{PRW~\cite{mou2020DDN}}& $\pm$ 0.025 & $\pm$ 0.448 &  $\pm$ 85.873& & $\pm$ 0.015 & $\pm$ 0.177 &  $\pm$ 18.588 \\
\multicolumn{1}{c|}{}&\multicolumn{1}{c}{} &  0.767 & 2.423 & 9.003 & &0.809 & 1.192 & \textbf{9.111} \\
\multicolumn{1}{c|}{}& \multirow{-2}{*}{${\text{G}_{\text{reco}_{\text{CCO}}}}$}& $\pm$ 0.023 & $\pm$ 0.582 &  $\pm$ 11.388& & $\pm$ 0.016 & $\pm$ 0.19 &  $\pm$ 7.606 \\
\multicolumn{1}{c|}{}& \multicolumn{1}{c}{} &  \textbf{0.768} & 2.332 &\textbf{ 6.609}  & &0.810 & 1.198 & 11.374\\
\multicolumn{1}{c|}{\multirow{-6}{*}{2D}}&\multirow{-2}{*}{${\text{G}_{\text{reco}_{\text{STARE}}}}$}& $\pm$ 0.023 & $\pm$ 0.533 &  $\pm$ 5.817 & &  $\pm$ 0.015 & $\pm$ 0.183 &  $\pm$ 9.095\\ 
\hline

\multicolumn{1}{c|}{}& p-values& $\thicksim 10^{-6}$  & 0.057 &  $\thicksim 10^{-6}$  & &  $0.832$ & $0.466$ &  $\thicksim 10^{-6}$ \\ 
\hline \hline
\multicolumn{1}{c|}{}&\multicolumn{1}{c}{} & 0.476 & \textbf{3.587} & 26.562& &  \textbf{0.756} & \textbf{1.488} & 3.203 \\
\multicolumn{1}{c|}{}&\multirow{-2}{*}{Segmentation} & $\pm$ 0.02 & $\pm$ 0.42 &  $\pm$ 10.264& & $\pm$ 0.015 & $\pm$ 0.211 &  $\pm$ 1.689\\
\multicolumn{1}{c|}{}&\multicolumn{1}{c}{} & \textbf{0.495} & 4.154 & \textbf{5.618} &  & 0.74 & 1.552 & \textbf{1.697}\\
\multicolumn{1}{c|}{\multirow{-4}{*}{3D}}& \multirow{-2}{*}{${\text{G}_{\text{reco}_{\text{IXI}}}}$}&$\pm$ 0.019 & $\pm$ 0.427 &  $\pm$ 2.411& & $\pm$ 0.014 & $\pm$ 0.21 &  $\pm$ 1.029\\ \hline
\multicolumn{1}{c|}{}& p-values& $\thicksim 10^{-4}$  & $\thicksim 10^{-6}$ &  $\thicksim 10^{-17}$  & &  $\thicksim 10^{-5}$ & $0.226$ &  $\thicksim 10^{-10}$ 
\end{tabular}}
\caption{Quantitative results obtained with our 2D and 3D reconnecting models on variational and deep learning segmentations. The p-values (from the t-test for normal distributions, or Wilcoxon test otherwise) are shown between the segmentation and ${\text{G}_{\text{reco}_{\text{STARE}}}}$ in 2D and the segmentation and ${\text{G}_{\text{reco}_{\text{IXI}}}}$ in 3D.}
\label{tab:chap3_variationnelle}
\end{table}

We observe that, in general, our post-processing either slightly increases the DSC and ASSD values of the segmentations or does not significantly change them (p-value $<$ 0.05). The slight decrease in the ASSD of the 3D variational approach and the DSC of the 3D deep learning approach can be attributed to some false reconnection of aligned artifacts.
The overall stability of the DSC and ASSD metrics is expected, because our post-processing primarily involves adding a few pixels to reconnect vessels as discussed before.
By contrast, we observe a drastic decrease in the connectivity metric $\epsilon_{\beta_0}$,  since our post-processing affects the connectivity of the segmentations. Specifically, we note a decrease of $>90\%$ in 2D, and $>80\%$ in 3D for the variational approach, and $>67\%$ in 2D, and $>47\%$ in 3D for the deep learning approach.

It is interesting to note that the model trained on the synthetic OpenCCO dataset yields slightly less inferior results, often due to the creation of false connections. Our reconnecting terms learn to reconnect only based on geometric features. Therefore, the closer the geometry of the vessels in the training dataset to that of the test dataset, the better the performance tends to be. Nonetheless, the drop of performance is quite small which makes our term very useful in a purely unsupervised context in 2D when no vascular annotation is available. In 3D, to the best of our knowledge, there is no synthetic vascular network generation software that yields vascular trees geometrically close enough to a real brain vascular network. In particular, the tortuosity of brain vessels is much higher than what is possible to generate with softwares like VascuSynth\footnote{https://vascusynth.cs.sfu.ca/Welcome.html} or OpenCCO. 


The PRW algorithm reconnects some vessels, as evidenced by the small decrease in $\epsilon_{\beta_0}$, but it is significantly less effective than our approach. PRW relies on probability maps to determine the direction of reconnection. However, we observed that the probability maps generated by both the variational and deep learning approaches exhibit a low dynamic range and are nearly binary. Consequently, the reconnection capability of PRW is limited on these types of segmentation results. In contrast, our approach directly relies on binary images to predict the reconnecting direction.

Additionally, qualitatively, PRW tends to produce more unrealistic reconnections, such as vessel enlargements, compared to those generated by our method.

\vspace{-0.2cm}
\section{Conclusion}
In this article, we introduced a novel vascular segmentation post-processing to favor vascular network connectivity. This post-processing can be used in an unsupervised or supervised context depending on the availability of vascular annotations on the dataset of interest. We conducted an extensive validation of our approach both in 2D and 3D and showed that our post-processing is robust to the size of disconnections, converges to a reconnected result when used iteratively, and significantly improves the connectivity of segmentation results.
We also compared our approach to a recent reconnecting post-processing algorithm, demonstrating that our method reconnects significantly more vessels and does so in a more realistic manner.
However, our approach is purely based on the vessel's geometric properties in binary segmentations and thus some false reconnections may appear. Future works include taking into account the intensities of the underlying image to avoid these false reconnections.


\section{Acknowledgments}


 This work was supported by Agence Nationale de la Recherche (ANR-22-CE45-0018, ANR-18-CE45-0018), LABEX PRIMES (ANR-11-LABX-0063). This work was granted access to the HPC resources of IDRIS under the allocations 2022-AD011013887 and 2023-AD011014452 made by GENCI.

\bibliographystyle{abbrv}
\bibliography{workshop_MICCAI}

\begin{thebibliography}{10}

\bibitem{bullitt2005vessel}
E.~Bullitt et~al.
\newblock Vessel tortuosity and brain tumor malignancy: {A} blinded study.
\newblock {\em Academic Radiology}, 2005.

\bibitem{esteves4573124plug}
S.~Carneiro-Esteves et~al.
\newblock A plug-and-play framework for curvilinear structure segmentation based on a learned reconnecting regularization.
\newblock {\em Neurocomputing}, 2024.

\bibitem{carrillo2007recursive}
J.~F. Carrillo et~al.
\newblock Recursive tracking of vascular tree axes in 3d medical images.
\newblock {\em International Journal of Computer Assisted Radiology and Surgery}, 1:331--339, 2007.

\bibitem{chan2006algorithms}
T.~Chan et~al.
\newblock Algorithms for finding global minimizers of image segmentation and denoising models.
\newblock {\em SIAM journal on applied mathematics}, 66(5):1632--1648, 2006.

\bibitem{chung2018accurate}
M.~Chung et~al.
\newblock Accurate liver vessel segmentation via active contour model with dense vessel candidates.
\newblock {\em Computer methods and programs in biomedicine}, 166:61--75, 2018.

\bibitem{clough2020topological}
J.~R. Clough et~al.
\newblock A topological loss function for deep-learning based image segmentation using persistent homology.
\newblock {\em IEEE transactions on pattern analysis and machine intelligence}, 44(12):8766--8778, 2020.

\bibitem{du2023retinal}
H.~Du et~al.
\newblock Retinal blood vessel segmentation by using the ms-lsdnet network and geometric skeleton reconnection method.
\newblock {\em Computers in Biology and Medicine}, 153:106416, 2023.

\bibitem{frangi1998multiscale}
A.~F. Frangi et~al.
\newblock Multiscale vessel enhancement filtering.
\newblock In {\em Medical Image Computing and Computer-Assisted Intervention—MICCAI’98: First International Conference Cambridge, MA, USA, October 11--13, 1998 Proceedings 1}, pages 130--137. Springer, 1998.

\bibitem{hakim2021regularizer}
L.~Hakim et~al.
\newblock Regularizer based on euler characteristic for retinal blood vessel segmentation.
\newblock {\em Pattern Recognition Letters}, 149:83--90, 2021.

\bibitem{hoover2000locating}
A.~Hoover et~al.
\newblock Locating blood vessels in retinal images by piecewise threshold probing of a matched filter response.
\newblock {\em IEEE Transactions on Medical imaging}, 19(3):203--210, 2000.

\bibitem{hu2019topology}
X.~Hu et~al.
\newblock Topology-preserving deep image segmentation.
\newblock {\em Advances in neural information processing systems}, 32, 2019.

\bibitem{joshi2011identification}
V.~S. Joshi et~al.
\newblock Identification and reconnection of interrupted vessels in retinal vessel segmentation.
\newblock In {\em 2011 IEEE International Symposium on Biomedical Imaging: From Nano to Macro}, pages 1416--1420. IEEE, 2011.

\bibitem{OpenCCO}
B.~Kerautret et~al.
\newblock {OpenCCO: An Implementation of Constrained Constructive Optimization for Generating 2D and 3D Vascular Trees}.
\newblock {\em {Image Processing On Line}}, 13:258--279, 2023.
\newblock \url{https://doi.org/10.5201/ipol.2023.477}.

\bibitem{kerfoot2018left}
E.~Kerfoot et~al.
\newblock Left-ventricle quantification using residual u-net.
\newblock In {\em International Workshop on Statistical Atlases and Computational Models of the Heart}, pages 371--380, 2018.

\bibitem{keshwani2020topnet}
D.~Keshwani et~al.
\newblock Topnet: Topology preserving metric learning for vessel tree reconstruction and labelling.
\newblock In {\em Medical Image Computing and Computer Assisted Intervention--MICCAI 2020: 23rd International Conference, Lima, Peru, October 4--8, 2020, Proceedings, Part VI 23}, pages 14--23. Springer, 2020.

\bibitem{jonas_article}
J.~Lamy et~al.
\newblock A benchmark framework for multi-region analysis of vesselness filters.
\newblock {\em IEEE Transactions on Medical Imaging}, PP:1--1, 07 2022.

\bibitem{li2023robust}
L.~Li et~al.
\newblock Robust segmentation via topology violation detection and feature synthesis.
\newblock In {\em International Conference on Medical Image Computing and Computer-Assisted Intervention}, pages 67--77. Springer, 2023.

\bibitem{liao2017progressive}
W.~Liao et~al.
\newblock Progressive minimal path method for segmentation of 2d and 3d line structures.
\newblock {\em IEEE transactions on pattern analysis and machine intelligence}, 40(3):696--709, 2017.

\bibitem{lin2023dtu}
M.~Lin et~al.
\newblock Dtu-net: Learning topological similarity for curvilinear structure segmentation.
\newblock In {\em International Conference on Information Processing in Medical Imaging}, pages 654--666. Springer, 2023.

\bibitem{Merveille:PAMI:17}
O.~Merveille et~al.
\newblock Curvilinear structure analysis by ranking the orientation responses of path operators.
\newblock {\em IEEE Transactions on Pattern Analysis and Machine Intelligence (PAMI)}, 40(2):304--317, 2018.

\bibitem{merveille2019n}
O.~Merveille et~al.
\newblock $n$ d variational restoration of curvilinear structures with prior-based directional regularization.
\newblock {\em IEEE Transactions on Image Processing}, 28(8):3848--3859, 2019.

\bibitem{miraucourt2015variational}
O.~Miraucourt et~al.
\newblock Variational method combined with frangi vesselness for tubular object segmentation.
\newblock In {\em Computational \& Mathematical Biomedical Engineering (CMBE)}, pages 485--488, 2015.

\bibitem{mou2020DDN}
L.~Mou et~al.
\newblock Dense dilated network with probability regularized walk for vessel detection.
\newblock {\em IEEE Transactions on Medical Imaging}, 39(5):1392--1403, 2020.

\bibitem{mou2021cs2}
L.~Mou et~al.
\newblock Cs2-net: Deep learning segmentation of curvilinear structures in medical imaging.
\newblock {\em Medical image analysis}, 67:101874, 2021.

\bibitem{peng2023curvilinear}
Y.~a. Peng et~al.
\newblock Curvilinear object segmentation in medical images based on odos filter and deep learning network.
\newblock {\em arXiv preprint arXiv:2301.07475}, 2023.

\bibitem{qiu2023corsegrec}
Y.~Qiu et~al.
\newblock Corsegrec: a topology-preserving scheme for extracting fully-connected coronary arteries from ct angiography.
\newblock In {\em International Conference on Medical Image Computing and Computer-Assisted Intervention}, pages 670--680. Springer, 2023.

\bibitem{rougé2023cascaded}
P.~Rougé et~al.
\newblock Cascaded multitask u-net using topological loss for vessel segmentation and centerline extraction, 2023.

\bibitem{rudin1992nonlinear}
L.~I. Rudin et~al.
\newblock Nonlinear total variation based noise removal algorithms.
\newblock {\em Physica D: nonlinear phenomena}, 60(1-4):259--268, 1992.

\bibitem{sanchesa2019cerebrovascular}
P.~Sanchesa et~al.
\newblock Cerebrovascular network segmentation of mra images with deep learning.
\newblock In {\em 2019 IEEE 16th international symposium on biomedical imaging (ISBI 2019)}, pages 768--771. IEEE, 2019.

\bibitem{sato1998three}
Y.~Sato et~al.
\newblock Three-dimensional multi-scale line filter for segmentation and visualization of curvilinear structures in medical images.
\newblock {\em Medical image analysis}, 2(2):143--168, 1998.

\bibitem{shi2022local}
T.~Shi et~al.
\newblock Local intensity order transformation for robust curvilinear object segmentation.
\newblock {\em IEEE Transactions on Image Processing}, 31:2557--2569, 2022.

\bibitem{shit2021cldice}
S.~Shit et~al.
\newblock cldice-a novel topology-preserving loss function for tubular structure segmentation.
\newblock In {\em Proceedings of the IEEE/CVF Conference on Computer Vision and Pattern Recognition}, pages 16560--16569, 2021.

\bibitem{staal2004ridge}
J.~Staal et~al.
\newblock Ridge-based vessel segmentation in color images of the retina.
\newblock {\em IEEE transactions on medical imaging}, 23(4):501--509, 2004.

\bibitem{tetteh2020deepvesselnet}
G.~Tetteh et~al.
\newblock Deepvesselnet: Vessel segmentation, centerline prediction, and bifurcation detection in 3-d angiographic volumes.
\newblock {\em Frontiers in Neuroscience}, page 1285, 2020.

\bibitem{Valderrama2023}
N.~Valderrama et~al.
\newblock Job-vs: Joint brain-vessel segmentation in tof-mra images.
\newblock In {\em ISBI 2023, IEEE International Symposium on Biomedical Imaging, 18-21 April 2023, Cartagena de Indias, Colombia}, Cartagena de Indias, 2023.

\bibitem{vaswani2017attention}
A.~Vaswani et~al.
\newblock Attention is all you need.
\newblock {\em Advances in neural information processing systems}, 30, 2017.

\bibitem{zhang2018reconnection}
J.~Zhang et~al.
\newblock Reconnection of interrupted curvilinear structures via cortically inspired completion for ophthalmologic images.
\newblock {\em IEEE Transactions on Biomedical Engineering}, 65(5):1151--1165, 2018.

\end{thebibliography}

\end{document}